\documentclass{jfm}
\usepackage{graphicx}
\usepackage{epstopdf,epsfig}

\shorttitle{Drag reduction in boiling TC turbulence}
\shortauthor{R. Ezeta et al.}

\title{Drag reduction in boiling Taylor-Couette turbulence}

\author{
Rodrigo Ezeta\aff{1},
Dennis Bakhuis\aff{1},
Sander G. Huisman\aff{1},
Chao Sun\aff{2,1},
\and Detlef Lohse\aff{1,3}
\corresp{\email{d.lohse@utwente.nl}},
}

\affiliation{
\aff{1}Physics of Fluids Group, Max Planck UT Center for Complex Fluid Dynamics,\break
MESA+ Institute and J.M. Burgers Centre for Fluid Dynamics, University of Twente, P.O. Box 217, 7500 AE Enschede, The Netherlands
\aff{2}Center for Combustion Energy, Key Laboratory for Thermal Science and Power Engineering of Ministry of Education, Department of Energy, Tsinghua University, Beijing, China
\aff{3}Max Planck Institute for Dynamics and Self-Organization, Am Fa{\ss}berg 17, G\"ottingen, Germany
}

\usepackage{setspace}
\usepackage{siunitx} 
\newcommand{\eref}[1]{equation \ref{#1}}
\newcommand{\fref}[1]{figure \ref{#1}}

\newcommand{\Fsref}[1]{Figures \ref{#1}}
\newcommand{\fsref}[1]{figures \ref{#1}}

\usepackage{tikz}
        
\newcommand{\red}[1]{\textcolor{red}{#1}}
\renewcommand{\red}[1]{#1}

\begin{document}

\maketitle

\begin{abstract}
We create a highly controlled lab environment---accessible to both global and local monitoring---to analyse turbulent boiling flows and in particular their  shear stress in a statistically  stationary state. Namely, by precisely monitoring the drag of strongly turbulent Taylor-Couette flow (the flow in between two co-axially rotating cylinders, Reynolds number $\textrm{Re}\approx 10^6$) during its transition from non-boiling to boiling, we show that the intuitive expectation, namely that a few volume percent of vapor bubbles would correspondingly change the global drag by a few percent, is wrong. Rather, we find that for these conditions  a dramatic global drag reduction of up to 45\% occurs. We connect this global result to our local
observations, showing that for major drag reduction 
the vapor bubble deformability is crucial, corresponding to Weber numbers larger than one. We compare our findings with those for turbulent flows with gas bubbles, which obey very different physics than vapor bubbles. Nonetheless, we  find remarkable similarities and explain these.
\end{abstract}


\section{Introduction}

When the temperature of a liquid increases, the corresponding vapor pressure $P_v$ also increases. Once the vapor pressure equals the surrounding pressure $P_{atm}$, all of the sudden {\it boiling} starts and vapor bubbles can grow \citep{prosperetti2017,brennen1995,dhir1998,theofanous2002,theofanous2002b,dhir2005,nikolayev2006,kim2009} (preferentially starting on nucleation sites such as on immersed microparticles), dramatically changing the characteristics of the flow \citep{gungor1986,weisman1983,tong2018,amalfi2016}. Similarly, {\it cavitation} requires that the pressure  is locally lowered so that it matches the vapor pressure at the given  temperature, and again vapor bubbles emerge, again dramatically changing the flow characteristics \citep{brennen1995,arn02}. Both boiling and cavitation  lie  at the basis of a vast range of different phenomena in daily life, nature, technology, and industrial processes. For boiling, many of them are  connected with energy conversion such as handling 
of LNGs (e.g., pumping through pipes or pipelines), liquified CO$_2$,  riser tubes of steam generators, boiler  tubes of power plants, or cooling channels of boiling water nuclear reactors. The sudden change of the global flow characteristics at the  boiling point can lead to catastrophic events \citep{gungor1986,weisman1983,tong2018,amalfi2016}. In nature, examples for such catastrophic events in boiling turbulent flows include volcano or geyser eruptions \citep{manga2006,toramaru2013}. For cavitation, one example is the cavitation behind rapidly rotating ship propellors, drastically reducing the propulsion \citep{carlton2012,geertsma2017} and another one is the occurrence of cavitating bubbles in turbomachinery, where the corresponding huge pressure fluctuations  can cause major damage \citep{dagostino2007}. In spite of the  relevance and the  omnipresence of boiling and cavitation in turbulent flows, 
the physics  governing the corresponding pressure drop and thus the reduced 
wall shear stress or propulsion is  not fully understood, also due to the lack of controlled experiments. 

\begin{figure}
\begin{center}
\includegraphics[width=\textwidth]{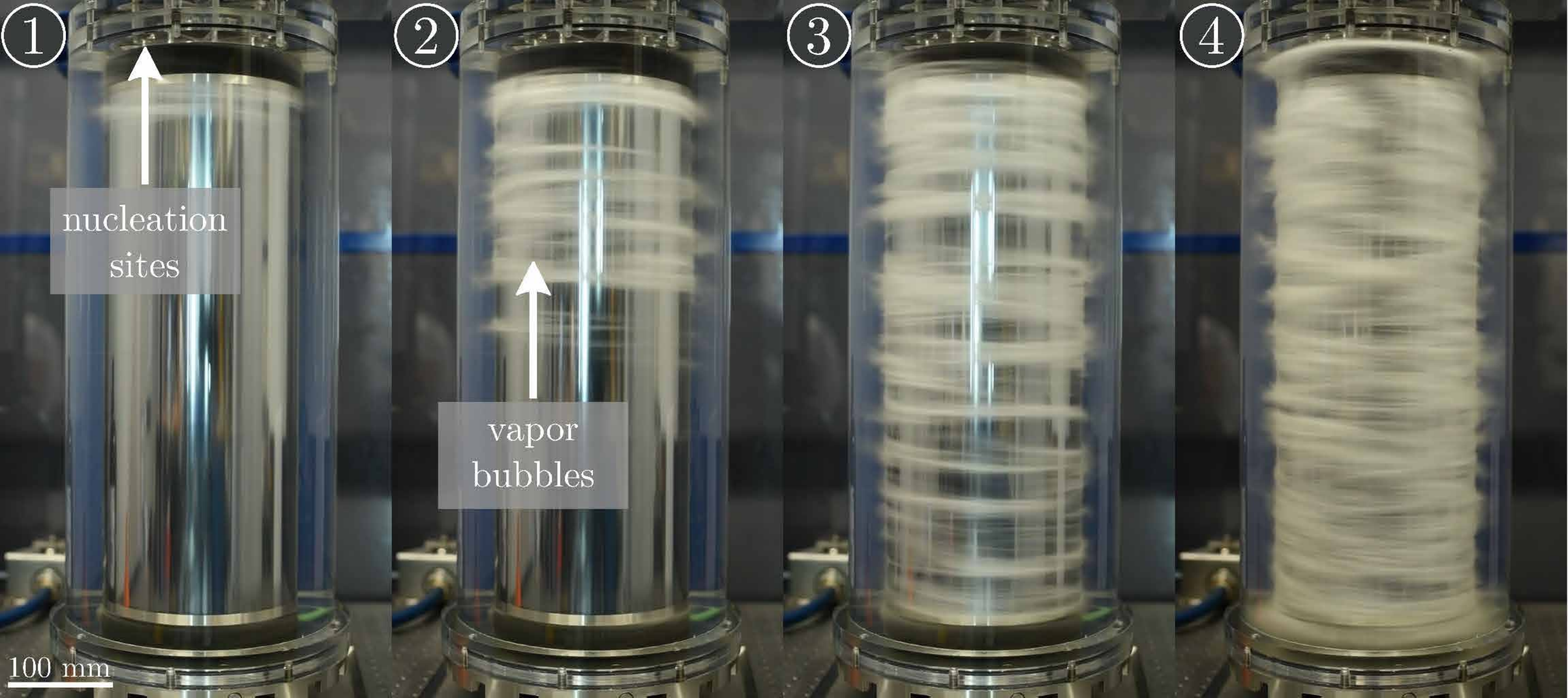} 
\caption{Vapor bubble generation and migration close to the surface of the IC in the boiling regime. Each image corresponds to a different stage during the experiment. Time moves from left to right: the numbers on top of each image corresponds  to the time steps shown in \fref{fig:figure2}d. The nucleation of vapor bubbles starts on top of the cell because the hydrostatic pressure there is smaller. The vapor bubble front then travels downwards by bubble dispersion until the surface of the IC is fully covered. Note that the volume fraction is increasing with time.}
\label{fig:figure1}
\end{center}
\end{figure}

\red{In this paper, we want to contribute to the understanding of boiling turbulent flows, by performing and analysing controlled boiling experiments in turbulent Taylor-Couette (TC) flow, i.e. the flow in between two coaxial co- or counterrotating cylinders. For a detailed overview of single-phase TC flow, we refer the reader to \citet{Grossmann2016}}. This flow has the unique advantage of being closed, with global balances, and no spatial transients. Also the underlying equations and boundary conditions are well-known and well-defined so that the turbulent flow is mathematically and numerically accessible. More concretely, the experiments are performed in the Boiling Twente Taylor-Couette 
(BTTC) facility \citep{Huisman2015}, which allows us to control and access at the same time both global and local flow quantities, namely to measure the torque $\mathcal{T}$ required to drive the cylinders at constant speed, the liquid temperature $T_{TC}$ in the cell, the pressure $P$ of the system, the volume fraction  $\alpha$ of the vapor, and in the beginning of the vapor bubble nucleation process even  the positions and sizes of individual vapor bubbles, all  as function of time, so that we can study the dynamics and evolution of the boiling process. The  focus of the work is on  the onset and evolution  of turbulent drag reduction induced by the nucleating vapor bubbles in the turbulent flow. We will  compare it with the case of air bubbles  in turbulent  flow with similar turbulence level and fixed gas volume fraction. For air bubbles, \citet{vangils2013} have shown that with a small volume fraction of only $\alpha \approx 4\%$,  very large drag reduction of $\approx 40\%$ can be achieved, provided the turbulence level is high, i.e. $\mathcal{O}(\text{Re})\approx 10^{6}$. We will not only compare the drag reduction effect of vapor and gas bubbles, but also their deformabilities, which have turned out to be essential for bubbly drag reduction \citep{lu2005,ber05,vangils2013,verschoof2016,spandan2018}.
  
\begin{figure}
\begin{center}
\includegraphics[scale=0.5]{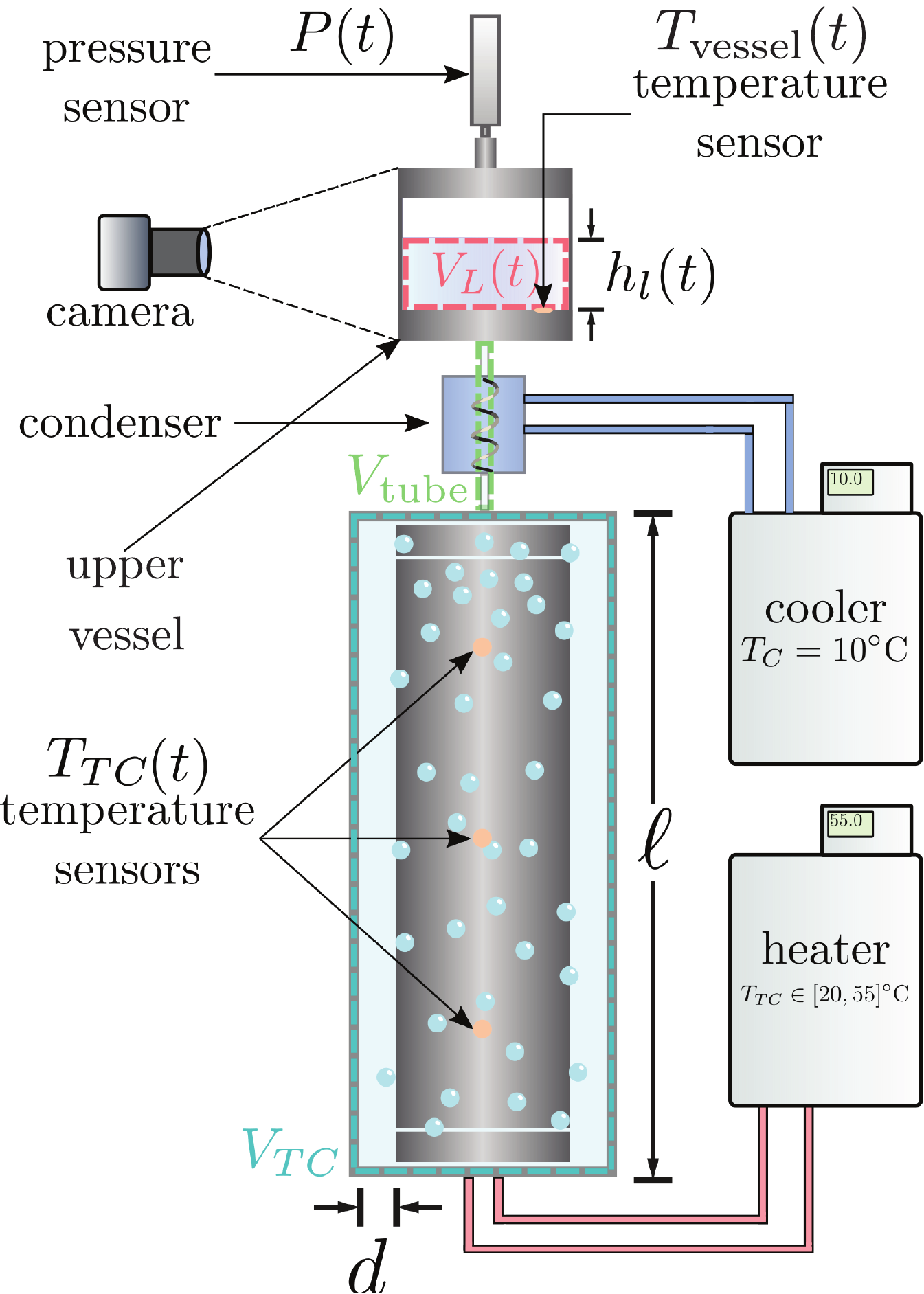}
\end{center}
\caption{Diagram of the experimental apparatus. The control volume defined by the blue dashed lines corresponds to the volume of the cell $V_{TC}$. Correspondingly, in green dashed lines we highlight the volume of the tubing $V_{\text{tube}}$. Finally, in red dashed lines we highlight the volume of the liquid height in the upper vessel $V_L(t)$. }
\label{fig:figure5}
\end{figure}

\begin{figure}
\begin{center}
\includegraphics[scale=0.4]{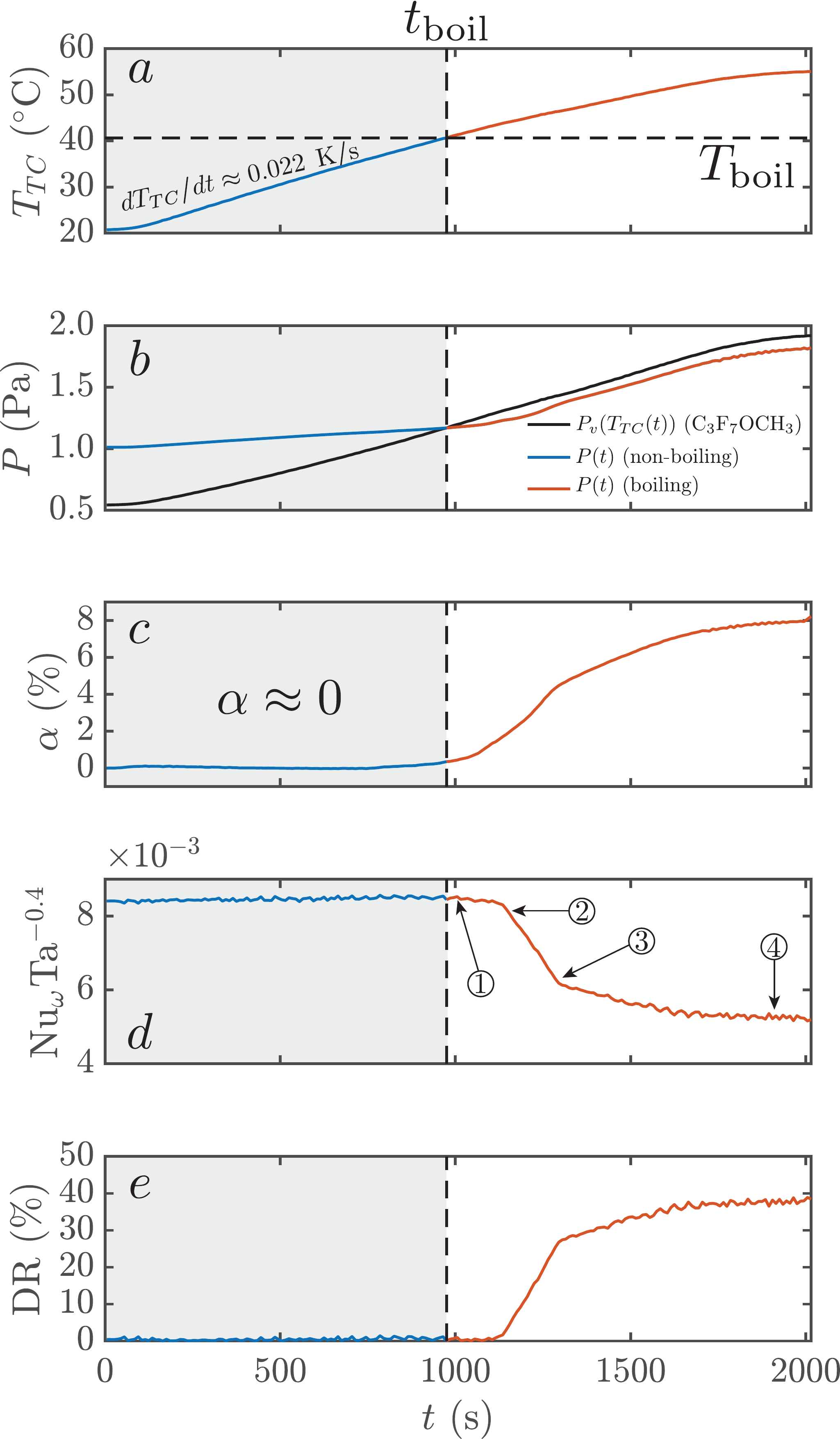}
\end{center}
\caption{(a) Liquid temperature $T_{TC}$, (b) pressure $P$, \red{c)} volume fraction $\alpha$, (d) the compensated Nusselt number $\text{Nu}_\omega\text{Ta}^{-0.4}$, and (e) the drag reduction (DR) as a function of time. The gray shaded area and the blue lines correspond to data in the non-boiling regime, i.e. $t<t_{\text{boil}}$. The boiling point is defined using the intersection $P=P_v$ at a certain time as it is shown in (b). The time steps in (d) correspond to the photographs shown in \fref{fig:figure1}.}
\label{fig:figure2}
\end{figure}

\section{Experimental setup and procedure}
\label{sec:setup}
In order to properly measure the drag reduction and the vapor volume fraction $\alpha$ in the flow as function of time, we have extended the BTTC facility \citep{Huisman2015} for boiling TC flow, see \fref{fig:figure1} for  examples of vapor bubbles in the BTTC at different stages in time. In \fref{fig:figure5}, we show a sketch of the experimental setup. The radius of the inner and outer cylinder of the BTTC are $r_{i}=\SI{75}{\milli \meter}$ and $r_o=\SI{105}{\milli \meter}$, respectively. The gap is then $d=r_o-r_i=\SI{30}{\milli \meter}$ and the radius ratio is $\eta=r_i/r_o=5/7=0.714$, which is very close to $\eta=0.716$ of the Twente Turbulent Taylor-Couette facility \citep{vangils2011}. The height of the cylinders is $\ell=\SI{549}{\milli \meter}$, which gives an aspect ratio of $\Gamma=\ell/d=18.3$. In order to allow for liquid expansion due to changes in temperature, the cell is connected to a transparent closed cylindrical container which we refer to as the \textit{upper vessel}. The cell and the upper vessel are connected via plastic tubing and an aluminum heat exchanger which is in direct contact with a water liquid bath. The coil and the liquid bath forms a very efficient condenser that condenses liquid back into the cell. The cell of the BTTC, the upper vessel, and the tubing that connects them form a closed reservoir; neither liquid nor vapor can leak out of the system. 

A circulator \texttt{Julabo FP50-HL} unit controls the temperature of the water bath for the condenser. A \texttt{PT 100} temperature sensor in the upper vessel monitors the liquid temperature $T_{\text{vessel}}(t)$. Since the system is closed, the increment in temperature translates into an increment in relative pressure which we monitor with an \texttt{Omega PXM409-002BGI} pressure sensor. The pressure signal is then calculated as $P(t)=P_{atm}+p(t)$, where $p(t)$ is the measured relative pressure. In order to avoid a possible overpressure of the system, a \SI{1}{\bar} pressure release valve is connected to the upper vessel. We use a \texttt{Nikon D300} camera to record the liquid height in the upper vessel $h_L(t)$ during the experiment. From this measurement, the liquid volume in the upper vessel can be calculated as $V_{\text{vessel}}(t)=\pi (D/2)^2 h_L(t)$, where $D=\SI{100}{\milli \meter}$ is the inner diameter of the upper vessel. The volume $V_{vessel}(t)$ and temperature of the liquid $T_{vessel}(t)$ in the upper vessel, along with the temperature of the liquid in the cell $T_{TC}(t)$ are used to calculate the volume fraction $\alpha(t)$ during the experiments as described in detail in appendix A.

The heating of the liquid phase is done via the surface of the IC which is itself heated through channels, where hot water can flow due to a second circulator \texttt{Julabo FP50-HL}. The IC is made out of stainless steel. Three temperature sensors distributed along the vertical axis of the IC measure the liquid temperature during the experiment. We take the average of the three sensors to calculate the liquid temperature inside the cell $T_{TC}(t)$. A \texttt{Althen 01167-051} hollow flanged reaction torque transducer (located inside the IC) measures the torque $\mathcal{T}$ required to drive the IC at constant speed. In addition to the torque experiments, we perform local measurements of the size of the vapor bubbles using high-speed imaging. The recordings were done with a \texttt{Photron SA-X} high-speed camera, with a frame rate of 13,500 fps. The framing of the camera results in a viewing area of $\SI{43.5}{\milli \meter} \times \SI{43.5}{\milli \meter}$ mm and it was recorded at mid-height. The focus plane of the camera is located at $(r-r_i)/d\approx 0.9$, thus the imaged bubbles are very close to the OC. The higher density of vapor bubbles near the IC makes the detection of these bubbles less reliable than for bubbles close to the outer cylinder.   

Boiling water requires liquid temperatures of $\approx\SI{100}{\celsius}$, which turns out to be impractical in the BTTC facility: the glass transition temperature of PMMA, from which the outer cylinder is made of, can be as low as $\approx \SI{82}{\celsius}$. 
Therefore,  instead of water, we use the commercially available low boiling point Novec Engineered Fluid 7000 ($\text{C}_3\text{F}_7\text{OCH}_3$) liquid instead. This liquid boils at roughly $\SI{34}{\celsius}$ at atmospheric pressure. The density of the liquid $\rho_\ell$, its kinematic viscosity $\nu_\ell$, and its surface tension (air/liquid) $\sigma$ depend on temperature. These dependencies can be found in \citet{rausch2015}.
 
 The experimental procedure is as follows: we fill the system with the low-boiling point liquid until an initial liquid height can be seen in the upper vessel. \red{Next, we rotate the IC at a fixed rotation frequency $f_i=\SI{20}{\Hz}$ while maintaining an initial liquid temperature of $\SI{20}{\celsius}$. Here, the ratio of centrifugal to gravitational acceleration is $a_{centr}/g=(2\pi f_i)^2 r_i/g\approx 121$.} At this stage, the entire fluid is still liquid. Once the system reaches thermal equilibrium, we strongly increase the liquid temperature $T_{TC}$ in the cell within the range $T_{TC}\in[\SI{20}{\celsius},\SI{55}{\celsius}]$. Eventually, the boiling point of the liquid is reached and we observe the nucleation of small vapor bubbles on top of the cylinder as shown in the first panel of \fref{fig:figure1}. The nucleation of vapor bubbles starts on top of the cell because the hydrostatic pressure there is smaller. Since the nucleated vapor bubbles are small, they are carried away by the flow and are able to migrate downwards close to the surface of the IC as shown in the second panel of \fref{fig:figure1}. \red{Note that the effective Froude number of the vapor bubbles, i.e. the ratio of centrifugal to gravitational forces is $\text{Fr}= \sqrt{\tilde{\rho} u_i^2 /((\tilde{\rho}-1)g r_i)}\approx 0.63$, where $\tilde{\rho}=\rho_v/\rho_\ell$ is the density ratio and $\rho_v$ the vapor density. } The bubble migration continues until the bubble front reaches the bottom plate of the BTTC (third panel of \fref{fig:figure1}). A casual inspection of the experiment, indicates that the volume fraction of vapor increases with temperature (see the third and fourth panels of \fref{fig:figure1}). In summary, \fref{fig:figure1} shows a typical boiling experiment where we highlight four different stages of the experiment in which the bubble nucleation and migration can be fully appreciated.

In \fref{fig:figure2}a we show the temperature ramp of the liquid phase for the experiment shown in \fref{fig:figure1}, where a heating rate of $dT_{TC}/dt\approx 0.022 \ \text{K}/\text{s}$ was used. This value is calculated by applying a linear fit to the data in the range $\SI{60}{\second}<t<t_{\text{boil}}$, where $60$ s is the time at which the liquid temperature is observed to increase linearly with time, and $t_{\text{boil}}$ is the time at which the liquid starts boiling. Since the system is closed, the pressure $P$ of the system increases monotonically with temperature as can be seen in \fref{fig:figure2}b. The boiling temperature $T_{\text{boil}}$ is a function of $P$; thus, we calculate the boiling point by plotting the vapor pressure of the liquid $P_v$ along with $P$ as shown in \fref{fig:figure2}b. The boiling point occurs at $t=t_{\text{boil}}$, i.e. the instant of time where $P=P_v$, and as a consequence $T_{\text{boil}} \equiv T_{TC}(t_\text{boil})$. In this manner, we define the \textit{non-boiling} regime as the time interval during the experiment where the liquid experiences pure thermal expansion and no vapor is present, i.e. $t<t_{\text{boil}}$ (or equivalently $T_{TC}<T_{\text{boil}}$). This region is shaded in gray in \fref{fig:figure2}. Conversely, we define a \textit{boiling} regime, where $t\geq t_{\text{boil}}$ (or equivalently $T_{TC}\geq T_{\text{boil}}$).

As mentioned before, the amount of vapor is seen to increase as the temperature increases beyond the boiling point. This effect can be seen throughout the third and fourth panels of \fref{fig:figure1}, where one observes more and more bubbles as time goes by. From the conservation of mass, based on the measurements of the liquid temperature $T_{\text{vessel}}$ and the liquid height $h_L$ in the upper vessel (see \fref{fig:figure5}), we can compute the instantaneous volume fraction $\alpha(t)$. The results are shown in \fref{fig:figure2}c, where we observe, indeed, that the instantaneous volume fraction $\alpha(t)$ increases monotonically until a maximum value of $\approx 8\%$ is reached at the end of the temperature ramp. The details of the calculation of $\alpha(t)$ can be found in appendix A. This calculation is self-consistent: within the non-boiling regime $\alpha\approx 0$ and $\alpha>0$ in the boiling regime. We would like to point out that no free parameters are involved in the calculation of $\alpha(t)$. It is rather remarkable that our boiling experiments yield such large and well-controlled values of the vapor volume fraction.

 \red{In order to quantify the turbulence level of the experiments, we use the Taylor number which is defined as \citep{eckhardt2007b,Grossmann2016}:}
 
\begin{equation}
\label{eq:Ta}
\text{Ta}=\left( \frac{(1+\eta)^4d^2(r_i+r_o)^2(2\pi f_i)^2} {64\eta^2}\right)  \frac{1}{\nu(\alpha) ^2},
\end{equation}

\noindent with $\nu(\alpha)=\nu_\ell(1+5\alpha/2)$, i.e. including the so-called Einstein correction for the viscosity due to the presence of the dispersed phase \citep{einstein1906}, $r_i$ and $r_o$ the radii of the inner and outer cylinder of the BTTC, respectively, and $\eta=r_i/r_o$ the radius ratio. The Taylor number and the Reynolds number $\text{Re}_i=2\pi f_i r_i d /\nu $ are related by $\text{Ta}=\frac{(1+\eta)^6}{64\eta ^4}\text{Re}_i^2$.

The response of the system, the torque or equivalently the angular velocity transfer, can---in analogy to the heat transfer in  Rayleigh-B\'enard flow---best be expressed in dimensionless form as the so-called generalized Nusselt number \citep{eckhardt2007b}

\begin{equation}
\label{eq:nusselt}
\text{Nu}_\omega=\left(\frac{r_o^2-r_i^2}{8\pi^2\ell_{\text{eff}} r_i^2 r_o^2 f_i} \right)\frac{\mathcal{T}}{\rho(\alpha) \nu(\alpha)},
\end{equation}

\noindent where $\ell_{\text{eff}}=\SI{489}{\mm}$ is the effective length along the cylinder where the torque is measured, and $\rho(\alpha)=\rho_\ell (1-\alpha) + \rho_v \alpha$ is the effective density of the medium. In the ultimate regime of TC turbulence \citep{Grossmann2016}, where both the bulk and the boundary layers are turbulent ($\text{Ta}>3\times 10^8$), it was found that $\text{Nu}_\omega \propto \text{Ta}^{0.4}$ \citep{Paoletti2011,vangils2011b,Huisman2012,Huisman2014}. In our experiments, Ta is of the order $\mathcal{O}(10^{12})$ and thus the compensated quantity $\text{Nu}_\omega\text{Ta}^{-0.4}$ can be used as a measure to characterize  the amount of drag reduction. In \fref{fig:figure2}d, we show that indeed $\text{Nu}_\omega \propto \text{Ta}^{0.4}$ during the experiment, up to a dramatic drop some time after the boiling point is reached. In the absence of a dispersed phase (non-boiling regime), $\text{Nu}_\omega\text{Ta}^{-0.4}$ should remain constant because the driving strength (i.e.,  the angular velocity $2\pi r_i f_i$ of the inner cylinder) is fixed and all the temperature-dependent quantities are contained in both $\text{Nu}_\omega$ and \text{Ta} (see \fref{fig:figure2}d). \red{Then, in the boiling regime, the occurrence of the vapor bubbles modifies the local angular velocity flow near the IC, causing a dramatic drop in the transport of momentum (Nusselt number) \citep{vangils2011b}, which we directly observe as a drop in the global torque.} This  is the  signature of drag reduction  in the flow, shown in \fref{fig:figure2}e. Notice how the start of the drop in the compensated Nusselt number and the corresponding drag reduction occur at a time after the boiling point is reached. This is simply because a certain amount of time (determined by turbulent diffusion) is needed for the vapor bubbles to migrate downwards and cover the entire surface of the IC. In the context of bubbly-induced drag reduction flow, transients in the wall shear stress are known to exist. The numerical study of \cite{xu2002} has shown for example, that in channel flow ($\text{Re}=3000$) and for air bubbles injected along the streamwise direction ($\approx 8\%$), the transients of shear stress are a function of the bubble size, i.e. larger bubbles produce higher transients. In our experiments, however, the Reynolds number is 3 decades larger ($\mathcal{O}(10^6)$) and it is known that transients are minor and that the flow quickly adjusts to the cylinder speed because of the strong convection of momentum by the turbulence \citep{vangils2011}. In addition, boiling experiments were conducted with different heating rates, where no discernible differences in the DR results were observed.
In other words, the time-dependent DR shown in \fref{fig:figure2}e for $t_{\text{boil}}<t<1250$ (first, second, and third panels of \fref{fig:figure1}) is mainly a consequence of the vapor bubbles axially redistributing along the IC.

\begin{figure}
\begin{center}
\includegraphics[scale=0.38]{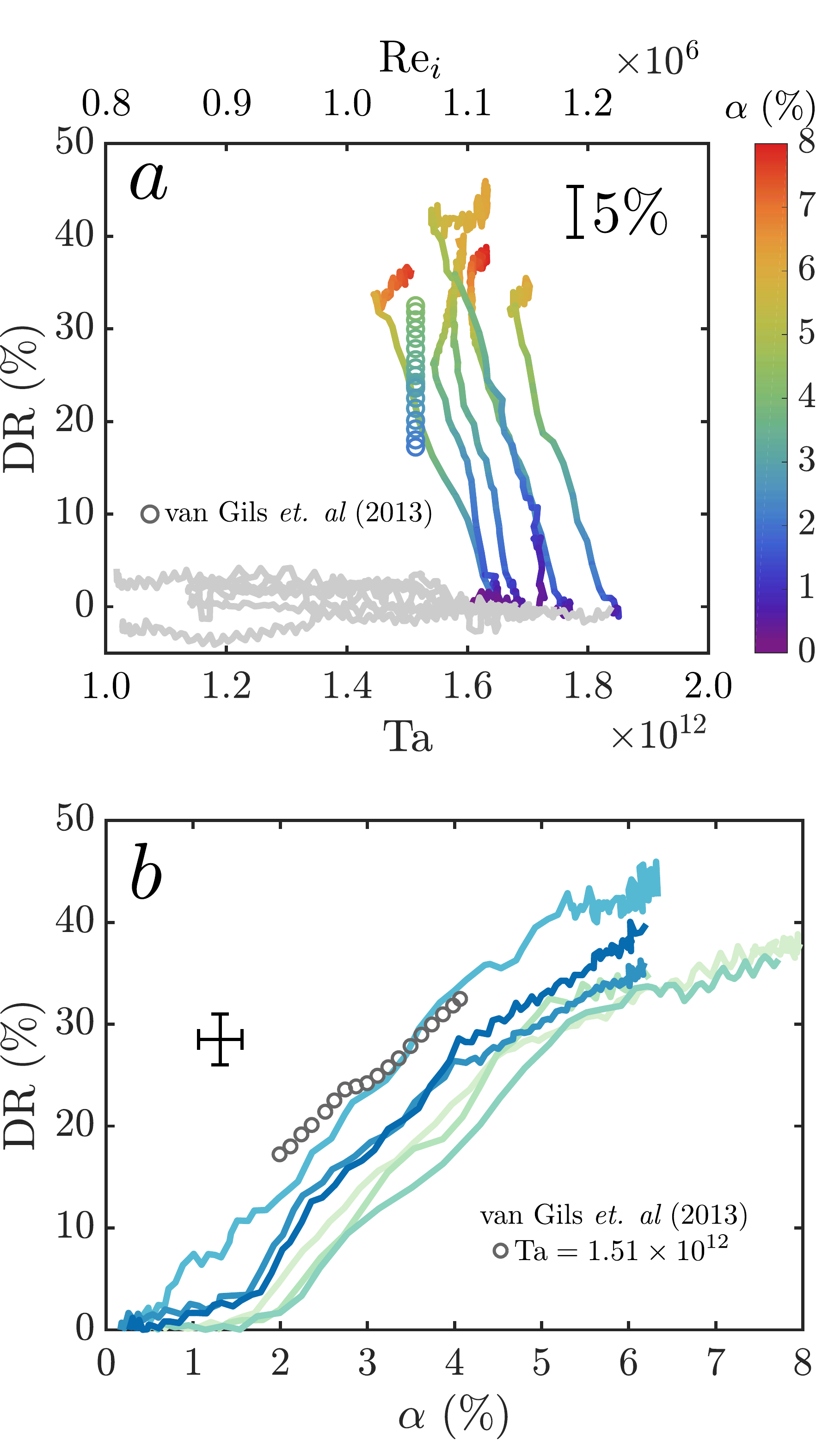}
\end{center}
\caption{(a) Drag reduction DR as a function of Ta and $\alpha$ for different experiments. Note that neither $\alpha$, nor DR, nor Ta are control parameters, but responses of the system to the temperature change. The horizontal axis on top of the figure represents the Reynolds number $\text{Re}_i$. The color bar represents the volume fraction $\alpha$. The gray data points correspond to data in the non-boiling regime ($\text{DR}\approx 0$, $\alpha\approx 0$). The open circles represent the drag reduction 
 obtained with air bubbles at a fixed $\text{Ta}=1.51\times 10^{12}$ \citep{vangils2013}. A 5\% error bar is shown that applies to all experiments. (b) DR as a function of the volume fraction $\alpha$. The colored lines represent the different experiments as shown in (a). The open circles correspond to the data of \citet{vangils2013} for drag reduction  using air bubbles. The error bars for both quantities DR (5\%) and $\alpha$ (0.5\%) are included. A 3D animation of the data is included in the supplementary material. Note the degree of reproducibility of our controlled experiments, which is remarkable for the boiling process, which is considered to be random and irregular.}
\label{fig:figure3}
\end{figure}

\section{Quantifying the drag reduction}
We quantify the level of drag reduction DR through

\begin{equation}
\label{eq:DR}
\textrm{DR}=1-\frac{\textrm{Nu}_\omega \textrm{Ta}^{-0.4}}{\textrm{Nu}_\omega \textrm{Ta}^{-0.4}(t=t_{\textrm{boil}})},
\end{equation}

\noindent where $\text{Nu}_\omega \text{Ta}^{-0.4}(t=t_{\text{boil}})$ is the compensated Nusselt number evaluated at the boiling point, where the system is still in a single phase state. In this way,  $\text{DR}>0$ corresponds to a finite amount of drag reduction (boiling regime), while $\text{DR}=0$ corresponds to zero drag reduction (non-boiling regime). We note that by introducing the correction for both the viscosity $\nu(\alpha)$ and density $\rho(\alpha)$ due to the presence of the dispersed phase, the net value of DR changes slightly as compared to the case when the correction is not used. In this study---as it also done in other studies of air bubble induced DR \citep{vangils2013,verschoof2016}---we set $\nu=\nu_\ell(1+5\alpha/2)$ and $\rho=\rho_\ell (1-\alpha)+\rho_v \alpha $ in order to draw accurate comparisons between our experiments with vapor and the case of air bubble injection. \red{Note that by introducing these corrections to the viscosity and density (via $\alpha$), the trivial effect of drag reduction due to the density decrease of the liquid-vapor mixture is already taken into account}. In \fref{fig:figure3}a, we show DR as function of Ta and $\alpha$ for the experiment described in \fsref{fig:figure1} and \ref{fig:figure2} along with five other experiments we performed, which confirm the  reproducibility of our controlled experiments. \Fsref{fig:figure3}ab  unambiguously reveal that drag reductions  approaching 45\% can be achieved in the boiling regime, with  vapor bubbles as dispersed phase, with a 
 volume fraction only up to 6--8\%. \red{An error bar of 5\% for DR and 0.5\% for $\alpha$ have been added to \fsref{fig:figure3}ab. These error bars were obtained from the repeatibility of the experiments.}
 
 Note  that  $\text{Ta}$ is a non-monotonic function in the boiling regime due to the vapor fraction  correction of the liquid viscosity. The reason is that when the boiling regime is reached, $\alpha$ increases at a faster rate than the rate at which $\nu_\ell$ decreases; and as a consequence the corrected viscosity $\nu$ increases which leads to a decrease of $\text{Ta}$ (see \eref{eq:Ta}). However, this effect is not significant. The converse takes place at the final stage of the experiment (see \fref{fig:figure2}c), when $\alpha$ saturates to a certain value and as a consequence the corrected viscosity $\nu$ decreases which leads to an increment of Ta (see \fref{fig:figure3}a). A 3D representation of \fref{fig:figure3} which shows the instantaneous drag reduction as a function of both Ta and $\alpha$ can be found in the supplementary movie section.

In \fref{fig:figure3}b we show the $\text{DR}$ as a function of the volume fraction $\alpha$, for all the corresponding experiments shown in \fref{fig:figure3}a. This figure reveals the amount of instantaneous
drag reduction  that can be achieved for given vapor fraction $\alpha$ at that instant. Again, it is remarkable that with only 6 - 8\% 
vapor bubble fraction drag reduction of nearly 45\% can be achieved. This resembles the large drag
reduction of up to 40\% achieved by the injection of only 4\% volume fraction of {\it air bubbles} 
into the TC system \citep{vangils2013}.

\section{Comparison to drag reduction with gas bubbles}
We will now compare in more detail, the drag reduction achieved in boiling turbulent flow (vapor bubbles) with the well-known effect of drag reduction in turbulent flow with gas bubbles \citep{mad84,mad85,merkle1992,deutsch2004,sug04,lu2005,ber05,san06,vanderberg2007,murai2008,ceccio2010,murai2014,elbing2013,vangils2013,kumagai2015,verschoof2016}. \red{Vapor and gas bubbles are fundamentally different \citep{prosperetti2017}. While the creation, growth, and stability of gas bubbles is entirely controlled by mass diffusion, vapor bubbles are controlled by the heat diffusion and by phase transitions. For typical flows, the ratio of the heat diffusion constant and the mass diffusion constant is $k_\ell/D_\ell = \mathcal{O}(100)$.} Since the surface tension in a vapor bubble is temperature-dependent, thermal Marangoni flows can further affect the vapor bubble dynamics. Given these major differences between vapor and gas bubbles, one wonders whether these major differences also reflect in different bubbly drag reduction behavior.  

The answer can be read off \fsref{fig:figure3}ab, in which we have also included the drag reduction data of \citet{vangils2013}, which correspond to the case of air bubbles at a fixed $\text{Ta}=1.51\times 10^{12}$. The  inspection of this figure reveals, strikingly, that both vapor and air bubbles produce a comparable amount of drag reduction  when $\alpha$ and $\text{Ta}$ are approximately the same for the two cases. \red{Note that in gas bubble injection experiments, the gas volume fraction $\alpha$ is a control parameter, whereas in the boiling experiments, the vapor volume fraction $\alpha$ is a response of the system to the temperature increase.} Indeed, for an equivalent Ta, the same amount of very large drag reduction can be obtained by either vapor or gas bubbles, given that their volume fraction is the same. 
 
\section{Bubble deformability}
\red{To achieve large drag reduction in high-\text{Re} flows with relative small gas bubble volume fraction, the gas bubble deformability has been identified as one of the crucial ingredients. This view is supported by experimental and numerical studies in high Reynolds number TC flows and other turbulent canonical flows: \citep{merkle1992,moriguchi2002,ber05,lu2005,shen2006,vanderberg2007,murai2008,vangils2013,murai2014,verschoof2016,rosenberg2016,spandan2018} Whether a bubble is deformable or not is determined by the corresponding Weber number which compares inertial and capillary forces, and it also influences the mobility of the bubbles. It is defined as}

\begin{equation}
\label{eq:weber}
\text{We}=\frac{\rho_\ell u_\theta^{\prime 2} d_b}{\sigma},
\end{equation}

 \noindent where $d_b$ is the typical bubble size, $u_\theta^{\prime 2}$ is the variance of the azimuthal velocity, $\sigma$ is the surface tension of the liquid/air interface. A large Weber number $\text{We} > 1 $ implies that the bubble is deformable. Indeed, \citet{verschoof2016} showed that for fixed gas volume fraction $\alpha \approx 4\%$  and fixed large Reynolds number, the large drag reduction ($\approx 40\%$) could basically be ``turned off'' by adding a surfactant, which hinders bubble coalescence and leads to much smaller bubbles with $\text{We} < 1$ in the strongly turbulent flow. 
 
 To obtain the Weber numbers of the vapor bubbles in our boiling experiments, we performed high-speed image recordings to measure the bubble size and shape. With this information at hand, we calculate the distribution of the Weber number for different volume fractions during the experiment. The velocity fluctuations  $u_\theta^{\prime}$ (required to calculate We) as function of Ta are  given by  $u_\theta^{\prime}=11.3\times 10^{-2}(\nu_\ell / d)\text{Ta}^{0.44}$, as measured earlier in the same BTTC facility \citep{ezeta2018}. The temperature variation during a high-speed measurement is $<\SI{0.3}{\kelvin}$. Therefore, all the values of the quantities in \eref{eq:weber} (except for $d_b$) are taken as the temperature-dependent  value at the beginning of each recording. 

 \begin{figure}
\begin{center}
\includegraphics[scale=0.38]{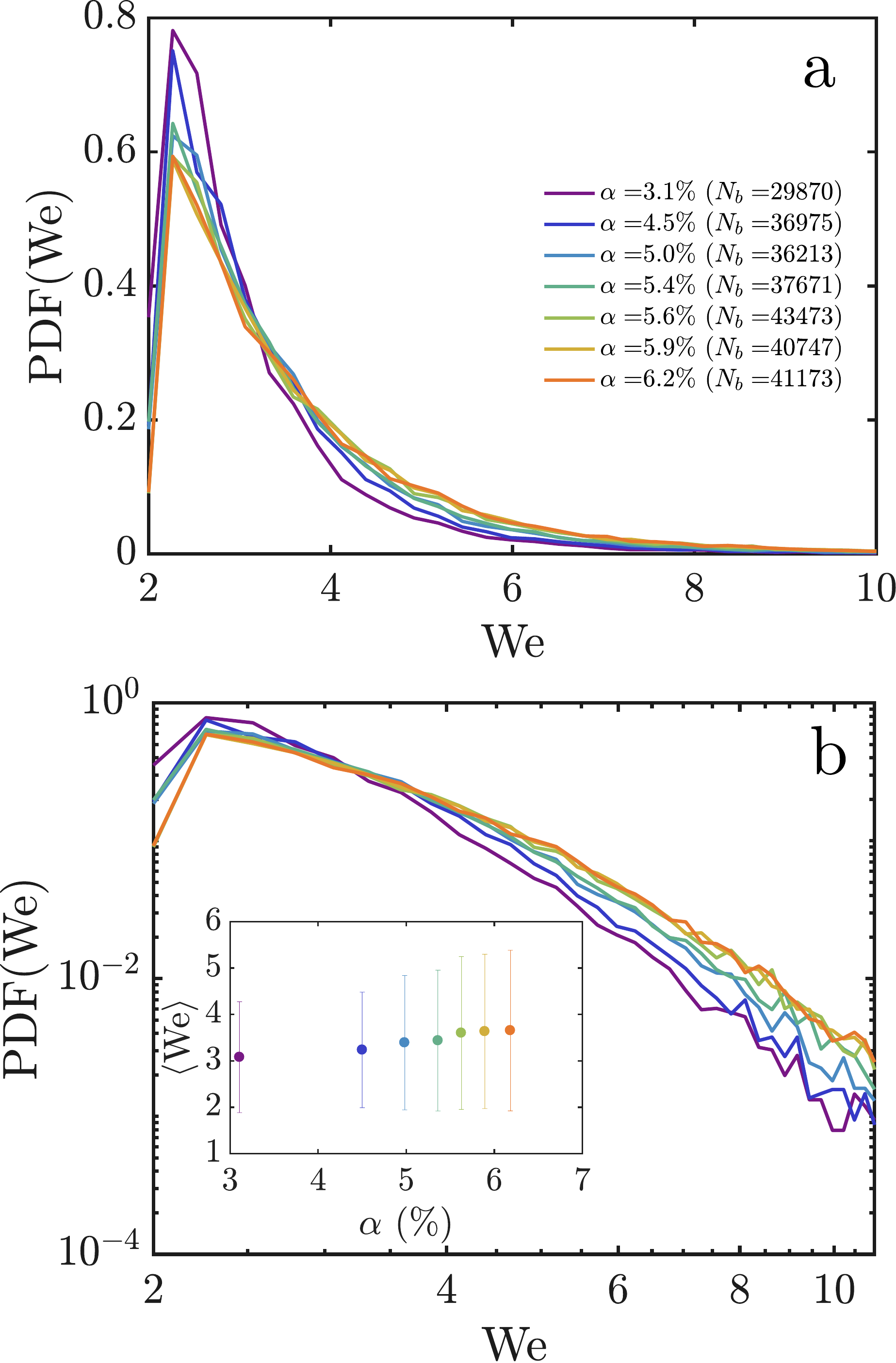}
\end{center}
\caption{(a) Probability density function of the bubble Weber number during the
 boiling experiment for different volume fractions in linear scale. The colors represent the variation of $\alpha$ as shown in the legend.
 $N_b$ is the number of detected bubbles in every measurement.  (b) Same as in (a) but in log-log scale. The inset in (b) represents the mean Weber number as function of the volume fraction. The error bars in (b) correspond to $\pm \sigma(\text{We})$, where $\sigma(\text{We})$ is the standard deviation of the Weber number for a given $\alpha$.  
 }
\label{fig:figure4}
\end{figure}

In \fref{fig:figure4}, we show the probability density function (PDF) and the mean value of the Weber number for different $\alpha$ during a typical experiment in the boiling regime. These distributions reveal that $\text{We}>1$, independent of the volume fraction, i.e. the vapor bubbles in our experiments are deformable, which supports the idea that also for vapor bubbles deformability is key for achieving large drag reduction, just  as shown for bubbly drag reduction with gas bubbles \citep{vangils2013,verschoof2016}. Moreover, we find that the maximum of all distributions lies at $\text{We}\approx 2.5$, which corresponds to bubbles of size $\approx 0.27 \ \text{mm}$. The tail of every PDF extends up to $\text{We}\approx 8$, which indicates that some of the bubbles are highly deformable. Furthermore, the inset in \fref{fig:figure4}b reveals that in the range of the volume fraction explored with the high-speed image experiments ($\alpha\in[3.1\%,6.2\%]$), the mean Weber number $\langle \text{We} \rangle$ slightly increases, namely from a value of $\langle \text{We} \rangle \approx 3.1$ to $\langle \text{We} \rangle \approx 3.7 $. Notice also that as the volume fraction increases, the probability to find a larger value of \text{We} is also larger for $\text{We}>3.5$, which indicates that the deformability of the vapor bubbles increases when the number of bubbles is increased. Since the variation of $\rho_\ell u_\theta^{\prime 2} / \sigma$ is very small throughout the boiling regime ($\approx 7\%$), this shows that with increasing volume fraction it is more likely to find bigger bubbles. This can also be seen at the other tail of the distribution for values of $\text{We}<3.5$, where the probability decreases with increasing volume fraction. So when advancing the boiling process, the emerging extra vapor manifests itself in larger vapor bubbles, be it by growth or coalescence, and not in more freshly nucleated small bubbles.

\section{Conclusions and Outlook}
We have investigated the transport properties (i.e., the drag) of strongly 
 turbulent boiling flow with increasing temperature in the highly  controlled BTTC setup and correlated
them  with the vapor bubble fraction and the vapor bubble characteristics. 
 Our highly reproducible and controlled findings reveal a sudden and dramatic drag reduction at the onset of boiling and that the emerging vapor bubbles are similarly 
  efficient in drag reduction 
  as injected air bubbles:  Nearly 45\% drag reduction
 can be achieved with a volume fraction of about  $\approx 6\%$. 
 In both cases the main reason for the drag reduction lies in the bubble deformability, reflected in 
 large We $>1$.  Furthermore, at later stages of the boiling process when the vapor fraction is higher, on average the bubbles are also bigger. 
 \red{This seems to be in line with our daily life experience when watching tea-water boil. However, this realization is not obvious  since the enhanced vapor fraction could also manifest itself in smaller, freshly-nucleated bubbles.}
 
 In our experiments (\fref{fig:figure2}a), the temperature increase is relatively modest, both in absolute numbers beyond the boiling temperature and in rate, due to experimental limitations. Furthermore, the drag reduction effect is spatially smeared out, as we measure the global drag of the whole cylinder. Nonetheless, within minutes the overall drag of the system reduces by a factor of 2. Within industrial devices, such as riser tubes of steam generators, boiler tubes of power plants, coolant channels of boiling water nuclear reactors, or when handling liquefied natural gases (LNGs) and liquefied CO$_2$, such sudden and large drag change can have dramatic consequences. In our experiments, the time scale of the sudden drag is determined by the heating rate and by the turbulent mixing of the emerging bubbles over the whole measurement volume. For larger heating rate and smaller volume, the rate of the drag change will be even more dramatic. Our experiments give guidelines on how to explore such events in a controlled and reproducible way.
 
\section*{Acknowledgments}
We would like to thank Sander Bonestroo for his valuable contribution to the bubble sizing experiments and Ruben Verschoof, Pim Bullee, and Andrea Prosperetti for various stimulating discussions. Also, we would like to thank Gert-Wim Bruggert, Bas Benschop, Martin Bos, Geert Mentink, Rindert Nauta, and Henk Waayer for their essential technical support. This work was funded by an ERC Advanced Grant, and by NWO-I and MCEC which are part of the Netherlands Organisation for Scientific Research (NWO). C. Sun acknowledges financial support from the Natural Science Foundation of China under Grant No.\ 11672156.

\section*{Supplementary movies}
Supplementary movies are available
 
\section*{Appendix A. Calculation of the volume fraction $\alpha$}
\label{sec:volume_fraction}
We calculate dynamically the volume fraction $\alpha(t)$ using the following conservation of mass argument. At the beginning of the experiment, i.e. the start of the temperature ramp, the temperature is $T(t_0)=T_0$. At this stage, the initial mass $m_0$ is composed only of the liquid mass which is known a priori. This is $m_0=\rho_\ell(T_0)(V_{TC}+V_{\text{tube}}+V_{\text{vessel}}(t=t_0))$, where $\rho_\ell$ is the liquid density of the $\text{C}_3\text{F}_7\text{OCH}_3$, $V_{TC}$ is the volume of the BTTC cell, $V_{\text{tube}}$ is the volume that corresponds to the tubing that connects the cell to the upper vessel and $V_{\text{vessel}}(t=t_0)$ is the liquid volume inside the upper vessel. Once the temperature ramp starts, the liquid experiences thermal expansion. This leads to a redistribution of the initial mass into $V_{TC}$, $V_{\text{tube}}$ and $V_{\text{vessel}}$ (see \fref{fig:figure5}):

\begin{eqnarray}
\label{eq:initial_mass}
m_0&=&m_{TC}\\  \nonumber
&+&\rho_\ell(T_{\text{vessel}}(t))V_{\text{tube}}+\rho_\ell(T_{\text{vessel}}(t))V_{\text{vessel}}(t),
\end{eqnarray}

\noindent where $m_{TC}$ is the mass inside the cell and $T_{\text{vessel}}(t)$ the temperature measured in the upper vessel, which we also assume is the temperature that corresponds to the tubing. Note that the only time dependent volume is $V_{\text{vessel}}(t)$.  As the temperature $T_{TC}$ increases, the boiling point $T_{boil}$ is eventually reached and vapor bubbles start nucleating. At this stage, the mass inside the cell $m_{TC}$ is a mixture of both the liquid and the vapor phase i.e.
\begin{equation}
\label{eq:mass_balance_mtc}
m_{TC}=\rho_\ell(T_{TC}(t))V_{\ell}(t)+\rho_v(T_{TC}(t),P(t))V_{v}(t),
\end{equation}
 where $T_{TC}$ is the measured temperature inside the cell, $V_\ell(t)$ is the volume occupied by the liquid inside the cell and $V_v(t)$ is the volume occupied by the vapor inside inside the cell, i.e. $V_\ell(t)+V_v(t)=V_{TC}$. The vapor density denoted by $\rho_v$ in \eref{eq:mass_balance_mtc} is dependent on both temperature and pressure due to the natural compressibility of the vapor phase. The vapor density is calculated by using tabulated values \citep{rausch2015} of $\rho_v(T,P_{\text{atm}})$ and assuming that the vapor experiences adiabatic expansion such that $\rho_v(T,P)=(P/P_{\text{atm}})\rho_v(T,P_{\text{atm}})$. In the absence of vapor, i.e. $\alpha=0$, $V_v(t<t_{boil})=0$. Using Eqs. \ref{eq:initial_mass} and \ref{eq:mass_balance_mtc}, along with the definition of the volume fraction $\alpha=V_v/V_{TC}$ and using that $1-\alpha=V_\ell/V_{TC}$, we find that

\begin{equation}
\label{eq:volume_fraction}
\alpha(t)=\frac{1-\frac{m_0}{\rho_\ell(T_{TC}(t)) V_{TC}}+\frac{\rho_\ell(T_{\text{vessel}}(t))}{\rho_\ell(T_{TC}(t))}\left( \frac{V_{\text{tube}}+V_{\text{vessel}}(t)}{V_{TC}} \right)}{1-\frac{\rho_v(T_{TC}(t),P(t))}{\rho_\ell(T_{TC}(t))}}.
\end{equation}

\bibliographystyle{jfm}
 Note the spaces between the initials
\bibliography{bibliography}

\end{document}